\documentclass[fleqn,usenatbib]{mnras}
\usepackage{newtxtext,newtxmath}
\usepackage[T1]{fontenc}
\usepackage{ae,aecompl}

\usepackage{graphicx}	% Including figure files
\usepackage{amsmath}	% Advanced maths commands

\usepackage{amssymb}	% Extra maths symbols
\usepackage{natbib}

\usepackage{upgreek}
\newcommand{\Msun}{M$_{\sun}$}
\newcommand{\orcid}[1]{\href{https://orcid.org/#1}{\includegraphics[width=10pt]{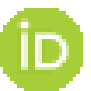}}}
\title[Chemical evolution of Galactic bulge]{The multi-zone chemical evolution of the Galactic bulge: predicting abundances for different radial zones}

\author[Cavichia et al.]{
O. Cavichia$^{1}$\orcid{0000-0002-7103-8036}\thanks{cavichia@unifei.edu.br} 
M. Moll{\'a}$^{2}$\orcid{0000-0003-0817-581X} and 
J.J. Baz{\'a}n$^{2}$\orcid{0000-0001-7699-3983}\\
$^{1}$ Instituto de F{\'i}sica e Qu{\'i}mica, Universidade Federal de Itajub{\'a}, Av. BPS, 1303, 37500-903, Itajub{\'a}-MG, Brazil\\
$^{2}$ Departamento de Investigaci\'{o}n B\'{a}sica, CIEMAT, Avda. Complutense 40. E-28040 Madrid, Spain.
}

\date{Last updated 2015 May 22; in original form 2013 September 5}

\pubyear{2022}

\begin{document}
\label{firstpage}
\pagerange{\pageref{firstpage}--\pageref{lastpage}}
\maketitle

% Abstract of the paper
\begin{abstract}
Due to its proximity, the stellar populations of the Galactic bulge (GB) can be resolved and can be studied in detail. This allows tracing the bulge metallicity distribution function (MDF) for different spatial regions within the bulge, which may give us clues about the bulge formation and evolution scenarios. In this work, we developed a chemical evolution model (CEM), taking into account the mass distribution in the bulge and disc, to derive the radial dependence of this time-scale in the Galaxy. Since the infall rate depends on that time scale in the CEM, the results of the model were used to test a scenario where the bulge is formed inside-out. The obtained results for the $[\alpha/\mbox{Fe}]$ {\sl vs.} [Fe/H] relationship, the MDF and the [Fe/H] radial gradient in the bulge have been compared to available data in the literature. The model is able to reproduce most of the observational data: the spread in the relation $[\alpha/\mbox{Fe}]$ {\sl vs.} [Fe/H], the MDF shape in different regions of the bulge, the [Fe/H] radial gradient inside it and the age-metallicity relation, as well as the [$\alpha$/Fe] evolution with age. The results of the model point to a scenario where the bulk of the bulge stars pre-existed the boxy/peanut X-shape bar formation. As a result, the classical origin of the GB is not ruled out and this scenario may be invoked to explain the chemical properties of the Galactic bulge.

\end{abstract}

% Select between one and six entries from the list of approved keywords.
% Don't make up new ones.

%%%%%%%%%%%%%%%%%%%%%%%%%%%%%%%%%%%%%%%%%%%%%%%%%%

%%%%%%%%%%%%%%%%% BODY OF PAPER %%%%%%%%%%%%%%%%%%

\section{Introduction}

The overdensities of light commonly found in the central regions of spiral galaxies are denominated bulges. These structures present a variety of stellar populations, metallicities and kinematics \citep[][and references therein]{fisher16}, which are probably related with their formation processes. Three main formation scenarios are currently considered in the literature to explain the bulges formation: 1) At higher redshifts, one can point out the dissipative collapse of a protogalactic gas cloud with a characteristic collapse time-scale given by the free-fall time-scale \citep{eggen62}; 2) Another formation process that may occur at higher redshifts is the accretion of substructures as disc material or external building blocks in a $\Lambda$CDM context \citep{immeli04, scannapieco03}; 3) At lower redsfhits, the secular formation from disc material through a bar formation \citep[e.g.][]{kormendy04} is also a possibility. In the first two scenarios, the outcome of the formation process is a classical bulge; in the last one is a boxy/peanut bulge. However, this dichotomy of bulge formation scenarios has been recently questioned, since observational evidence points to a joint formation and evolution of the bulge and disc \citep{breda18,breda20}.

Classical bulges are similar to elliptical galaxies in many aspects: they present low star formation rate (SFR), spheroidal shape with surface brightness profiles being well approximated by the \citet{sersic68} profile and a S\'ersic index $n \gtrapprox 3$, show stellar kinematics dominated by velocity dispersion \citep{kormendy04}, and correlate with scaling relations for normal elliptical galaxies \citep[][and references therein]{fisher10}. On the other hand, boxy/peanut bulges in late-type galaxies are generally best described by nearly exponential surface brightness profile, presenting S\'ersic functions with $n  \lessapprox 2$. Furthermore, they are dominated by a disc morphology such as the spiral structure, supported by coherent rotation, and are actively forming stars \citep{kormendy04}. 

Due to its proximity, the Milky Way bulge, or Galactic bulge, (hereinafter GB or bulge), is basically the only galaxy bulge that can be resolved and can be studied in greater detail. For many years, the GB was believed to present a single and wide metallicity distribution function (MDF) \citep{mcwilliam94,fulbright06}, which was consistent with a bulge composed mostly by old stars \citep{ortolani95} formed early by means of a quick dissipational collapse \citep{molla95,ballero07}. Later, observations \citep[e.g.][]{zoccali08} showed that the MDF of the GB was more complex, presenting asymmetries not detected before. They find a sharper cutoff on the high-metallicity side, and a MDF narrower than previously measured. They concluded proposing a scenario in which both infall and outflow were important during the bulge formation and, then, suggested the presence of a negative metallicity radial gradient. 

The first evidence for a bimodality in the bulge MDF was presented by \citet{hill11}. 
In turn, \citet{babusiaux10} analysed the relationship between chemical abundances and kinematics detecting that the metal-rich (MR) population has bar-like kinematics, while the metal-poor (MP) population shows kinematics corresponding to an old spheroid or a thick disc. More recently, \citet{zoccali17} have demonstrated that these two components also present a different spatial distribution. The MP dominates the MDF at high latitudes and the MR one dominates closer to the Galactic plane. However, at latitude smaller than $|b| = 3\degr$, the fraction of stars in the MP component starts to increase again. As a result of this mix of the two populations, a negative radial gradient is observed for the mean metallicity. This negative radial gradient was detected inside the GB in both vertical \citep{zoccali08} and radial directions \citep{ness13, fuentes14}.

Nowadays, the bulge MDF can also be traced for different spatial regions within the bulge, which  may give us clues about the bulge formation and evolution scenarios. Recent results point towards a bulge MDF having large variations across the bulge area, and where MP stars are very abundant in the inner few degrees of the Milky Way bulge \citep{zoccali18}. Although the MP component is centrally concentrated in absolute number \citep{zoccali17}, the relative number of MP to MR populations changes dramatically across the bulge minor axis (see figure 7 of these authors). However, only the radial gradient exhibited by the MP population may corresponds to an inherent gradient \citep{rojas-arrigada14} and can be interpreted as the signature of classical bulge formation.

In the last years, we have witnessed an impressive increasing amount of the available observational data of the GB thanks to the new spectroscopic surveys, as e.g. the Gaia-ESO \citep{gilmore12}, APOGEE \citep{majewski17}, ARGOS \citep{freeman13}, in addition to GIBS \citep{zoccali17}; and also photometric surveys, e.g. VVV survey \citep{minniti10}, Gaia mission \citep{brown21}, and PIGS \citep{arentsen20}. These surveys confirmed that the bulge MDF is clearly bimodal and that MP and MR stars have different spatial distributions and radial velocity dispersions. However, the currently available data is not sufficient to form a consensus in the literature regarding if they point to a GB formation scenario formed by a bar instability or formed by a spheroidal collapse and, depending on the observational data used, one or more scenario can be invoked to explain the observations, arriving this way at the so-called {\sl mixed scenario}. 

Additionally, even with the recent observational successes, the time-scales for formation and chemical enrichment of the GB remain a subject of debate. With regard to the $\alpha$-element abundances in the GB, \citet{johnson13} found that the relation [$\alpha/\mbox{Fe}$] {\sl vs.} $[\mbox{Fe}/\mbox{H}]$ is similar across the GB fields, indicating that the gas is well mixed and homogeneous, or, if there was any difference, it was erased during the GB evolution. 
On the other hand, \citet{hasselquist20} have recently presented stellar age distributions of the Milky Way bulge finding that the stars are predominately old, with half of the stars in their sample older than 8 Gyr. They also find that stars out of the plane are less enriched by a factor of three than stars in the plane, suggesting an inside-out bulge formation scenario where the stellar populations have not yet been fully mixed. This result is also supported by observations in other spiral galaxies, as e.g. by \citet{gonzalez-delgado14a,gonzalez-delgado15}. They have shown that, regardless of morphology, galaxies present a radial gradient of stellar ages, including their central regions, indicating that their stellar masses were assembled from inside-out. Besides that, MR and MP stars are in different locus in the [$\alpha$/Fe] {\it vs.} [Fe/H] plane. They both display a sequence like distribution, but recently, based on larger and more precise datasets, it has been found that even there is a gap (a low density region) between them, as detected by \citet{rojas-arriagada19} and, specially, \citet{queiroz21}.

Regarding this aspect, CEM are important tools to understand the formation and evolution of the components of MWG and other galaxies in the universe. The outputs of a CEM depend on the physics adopted, e.g. the stellar yields, initial mass function, star formation rate and gas infall law. The model results may be compared with the observational data, and the physics adopted may then be self-consistently changed until satisfactory results are achieved. In fact, one of the most important constraints for CEM is the MDF, which is sensible to the time-scale that the region was formed. Another widely used observational constraint to select the best physics as input of a CEM is the $[\alpha/\mbox{Fe}]$ relative abundances \citep{molla00,ballero07,cavichia14} since, as in the case of the MDF, this ratio is also dependent upon the time-scale involved in the formation of the corresponding region, one of the most important ingredients of models. 

The CEM from \citet{matteucci90} was the first model that predicted the bulge MDF. Their model mimics the {\it closed box} simple model, using a fast collapse time-scale and an IMF flatter than the standard \citet{salpeter55} towards the high-mass end. Later, \citet{molla95} developed a CEM to follow the bulge chemical evolution in a multi-zone approach, integrated with the evolution of the disc and halo, finding that the MDF, as observed at that time, was well reproduced with a normal IMF if the collapse time-scale was short.  \citet{ballero07} updated the model from \citet{matteucci90} and concluded that the bulge was formed by a fast time-scale and a flat IMF at higher stellar masses was again needed. More recently, \citet{grieco12,tsujimoto12,matteucci19} developed CEMs to explain the existence of the two main stellar populations observed in the bulge. \citet{haywood18} applied a modified closed-box model to the disc and bulge and proposed a disc origin for the bulge. \citet{fragkoudi18} also proposed a disc origin for the boxy bulge by modelling its formation through N-body simulations for a composite (thin+thick) stellar disc which evolves secularly to form a bar and a boxy/peanut bulge. In \citet{cavichia14}, the chemical evolution of the GB was simulated by the gravitational collapse of a spheroid connected to the halo and disc, including radial gas flows to explore the effects of the bar in the GB formation and evolution. In that work, it was demonstrated that the bar increases the star formation rate and modifies the Oxygen abundances, but only produces small variations on the MDF.

In this work we propose to use the bulge MDF and [$\alpha/\mbox{Fe}$] ratios to constrain the time-scale of the bulge formation in order to test a scenario where the bulge is formed inside-out in a multi-zone approach. From the observational side there is available data supporting this scenario, as e.g. in the recent work from \citet{hasselquist20}. Therefore, in light of the new available observational data derived from several surveys that observed the GB regions, it is a good opportunity to revisit the bulge formation and evolution scenarios. This paper is organized as follows: we describe the model in Section~\ref{model}, particularly the modifications done compared with our previous works, as the computation of the radial distribution of mass within the bulge and the new adopted stellar yields.  In Section~\ref{results}, we show our results related to the MDF and to the relative abundances [$\alpha/\mbox{Fe}$], as their variations in the different regions, and their comparison with the observational data. In Section~\ref{discussion}, we present a discussion of the results of this work in the context of our current understanding of the nature and origin of the Galactic bulge. Finally, our conclusions are summarized in Section~\ref{conclusions}.

\section{Model Description}
\label{model}
\subsection{Summary}

Our model is an update of the one from \citet{molla05,cavichia14}, in which the GB is formed assuming an inside-out growth scenario where the accretion of metal-poor gas from the halo builds the spheroidal component and also the disc. The disc is divided in concentric rings each one 1\,kpc wide. In this new work, the bulge is divided in concentric shells of 0.25\,kpc. The infall rate for each radial region is computed following prescriptions by \citet{molla16}, and it is obtained by imposing that after a Hubble time the system ends with a mass as observed:
\begin{equation}
\tau(R)=-\frac{13.2}{\ln{\left(1-\frac{\Delta M_{\textrm{\scriptsize D,B}}(R)}{\Delta M_{\textrm{\scriptsize tot}}(R)}\right)}}\,[\mbox{Gyr}].
\label{eq:tcoll}
\end{equation}
Here the characteristic time-scale, also referred as {\it collapse time-scale}, $\tau(R)$, is defined as the time necessary for the disc or bulge mass in each ring/shell, $\Delta M_{\textrm{\scriptsize D,B}}(R)$, reach the actual value at the present time through the prescribed infall formalism.  The total mass included in each cylindrical region above and below the corresponding radial region in which the gas will fall is defined as $\Delta
M_{\textrm{\scriptsize tot}}(R)=M_{\textrm{\scriptsize
tot}}(<R)-M_{\textrm{\scriptsize tot}}(<R-1)$, where: 
\begin{equation}
M_{\textrm{\scriptsize tot}}(R)=M_{\textrm{\scriptsize D}}(R)+M_{\textrm{\scriptsize H}}(R)+M_{\textrm{\scriptsize B}}(R),
\label{eq:mtot}
\end{equation}
where $M_{\textrm{\scriptsize D}}(R)$, $M_{\textrm{\scriptsize B}}(R)$ and $M_{\textrm{\scriptsize H}}(R)$ correspond to the disk, the bulge and the halo dynamical masses.  Following the prescriptions from \citet{salucci07}, we obtain the rotation curves for the halo and disc, and from them their corresponding radial mass distributions. This model was already applied successfully to the MWG disc in \citet{molla19a} and \citet{molla19b}, where the reader is referred to for more details. 

The adopted mass model is the one named 12.01, which corresponds to a Galaxy total mass $1.02\times10^{12}$\,\Msun\, and a disc mass $M_{\rm D}=7.20\times10^{10}$\,\Msun. The bulge mass $M_{\rm B}=1.93\times10^{10}$\,\Msun, which is in agreement with the most recent determinations \citep{valenti16,portail17}. The bulge effective radius from our models is $R_{\rm e}=0.536$\,kpc and the bulge radius $R_{\rm B}=2.1$\,kpc. We divided the GB in nine concentric regions, being the innermost a sphere of 125\,pc in radius and centred at 0\,kpc. The subsequent regions are spherical shells with external radii 0.25, 0.50 0.75, 1.25, 1.50, 1.75 and 2.00\,kpc, each one 0.25\,kpc wide. The bulge mass distribution is obtained allowing gas infalling with a infall law as in \citet{molla16}. In the innermost region, corresponding to $R=0$, we have included the mass of the supermassive black hole, but the mass of the nuclear stellar disc (NSD) is not considered in the model. However, the stellar mass of the nuclear bulge is estimated to be $1.4 \pm 0.6 \times 10^9 $\,\Msun \citep{launhardt02}, which is of the order of the value that we are estimating in this work for the central region at $R=0$ (see Table~\ref{tab:tcoll} below). Moreover, in the subsequent bulge regions, $R > 0$,  the mass of the disc is included as in \citet{molla16}.

The Initial Mass function (IMF) adopted is the one from \citet{kroupa01} in the range 0.15 -- 100\,\Msun\ for all regions of the Galaxy. This is justified, since there is no evidence for variations of the IMF in the MWG \citep[see a discussion by][and references therein]{barbuy18}. We have adopted two stellar mass intervals: 1) $0.15\, \le m<4$\,\Msun\ for low-mass stars; and  $ 4 \le m \le 100$\,\Msun\ for intermediate-mass and massive stars. The star formation (SF) process follows a Schmidt law in the halo regions. In the disc, the SF occurs in two steps: first, molecular clouds formed from diffuse gas; then, stars form through cloud-cloud collisions. Another possibility for the SF process is the interaction of massive stars with the molecular clouds surrounding them. More details about the SF process are given in \citet{molla05,molla16,molla19a}.

\subsection{Modelling the bulge mass profile}
\label{sec:maths} % used for referring to this section from elsewhere

In the case of spiral galaxies, the central bulge brightness profile can be well described by a S\'{e}rsic function \citep{sersic68}:
\begin{equation}
    I(R)=I_0\exp\left[-\left(\frac{R}{h}\right)^{1/n} \right],
    \label{eq:sersic}
\end{equation}
where $I_0$ is the central intensity, $h$ is a scale radius and $n$ is the S\'{e}rsic index. By assuming that the mass distribution follows the light distribution, the bulge total mass can be calculated by integrating Eq.~\ref{eq:sersic}, thus:
\begin{equation}
    M_B=2\pi I_0\int_0^\infty R\exp\left[-\left(\frac{R}{h}\right)^{1/n} \right]dR.
    \label{eq:mass}
\end{equation}

Performing the modification: $x=(R/h)^{1/n}$ in Eq.~(\ref{eq:mass}), we have:
\begin{equation}
    M_B=2\pi I_0 h^2 n \int_0^\infty x^{2n-1}e^{-x}dx,
    \label{eq:mass2}
\end{equation}
which is the definition of the gamma function \citep{arfken05}:
\begin{equation}
    \Gamma(2n)=\int_0^\infty t^{2n-1}e^{-x}dx=(2n-1)!,
    \label{eq:gamma}
\end{equation}
with $n$ being an integer. 
Therefore, the bulge total mass is:
\begin{equation}
    M_B=\pi I_0 h^2 (2n)!.
    \label{eq:mass3}
\end{equation}

In the same way, the bulge mass enclosed in a radius $R$ can be calculated as:
\begin{equation}
M_B(<R)=2\pi I_0h^2n \int_0^{\xi} x^{2n-1}e^{-x}dx,
\label{eq:mass_bul_r_int}
\end{equation}
where $\xi = (R/h)^{1/n}$. Using the property of the gamma function: $\gamma(2n,\xi)=\Gamma(2n)-\Gamma(2n,\xi)$, where $\gamma(2n,\xi)$ and $\Gamma(2n,\xi)$ are the lower and upper incomplete gamma functions, defined, respectively, as:
\begin{equation}
    \gamma(2n,\xi)=\int_0^\xi x^{2n-1}e^{-x}dx,
    \label{eq:gamma_lower}
\end{equation}
\begin{equation}
    \Gamma(2n,\xi)=\int_{\xi}^\infty x^{2n-1}e^{-x}dx,
    \label{eq:gamma_upper}
\end{equation} 
and taking $2n$ as a positive integer, Eq.~\ref{eq:mass_bul_r_int} turns out to be \citep{arfken05}:
\begin{equation}
    M_B(<R)=M_B \left(1-e^{-\xi}\sum_{i=0}^{2n-1}\frac{\xi^i}{i!}\right),
\label{eq:mass_bul_r}    
\end{equation}
where $M_B=\pi I_0h^2(2n)!$. 

The scale radius $h$ is obtained by 
using the definition of the effective radius $R_{\rm e}$, that is, the one that encloses half of the light of the model:
$I(R_{\rm e})=I_{\rm tot}/2$. So that:
\begin{equation}
    2\pi \int_0^{R_{\rm e}} R\,I(R)dR=\frac{2\pi}{2} \int_0^\infty R\,I(R)dR.
    \label{eq:mass_equality}
\end{equation}
Using again the lower incomplete gamma function $\gamma(2n,k)$, with $R_{\rm e}=k^nh$ and the properties  of the gamma function $2\gamma(2n,k)=\Gamma(2n)$ and $\gamma(2n,k)=\Gamma(2n)-\Gamma(2n,k)$, Eq.~\ref{eq:mass_equality} yields: 
\begin{equation}
\Gamma(2n)=2\Gamma(2n,k).
\label{eq:find_k}
\end{equation}
As $2n$ is a positive integer, we may write:
\begin{equation}
    \Gamma(2n,k)=(2n-1)!e^{-k}\sum_{i=0}^{2n-1}\frac{k^i}{i!}.
    \label{eq:gamma_k}
\end{equation}
Given the bulge effective radius $R_{\rm e}$, by solving Eq.~\ref{eq:find_k} by means of Eq.~\ref{eq:gamma_k}, we can obtain $h$ and $k$ for different values of the S\'{e}rsic index $n$. The bulge effective radius is obtained from our previous paper, \citet{molla16}, where we explored the observed correlations between disc and bulge structural parameters (see equation 11 of that paper). In this model, a Milky Way sized galaxy yields $R_{\rm e}=0.536$\,kpc, $M_{\rm B}=1.93\times10^{10}$~\Msun. Table \ref{tab:k_h_param} lists the values of the parameters $k$ and $h$ for different S\'{e}rsic indexes: $n=2$ (boxy/peanut bulge) and $n=4$ (classical bulge).
\begin{table}
	\centering
	\caption{Scale parameters for different S\'{e}rsic indexes.}
	\label{tab:k_h_param}
	\begin{tabular}{lcc} % four columns, alignment for each
		\hline
		$n$ & 2 & 4 \\
		\hline
		$k$ & 3.67206 & 7.66925\\
		$h$ (kpc) & 0.04308 & $1.67944\times 10^{-4}$\\
		\hline
	\end{tabular}
\end{table}

In the central region, ($R = 0$) the mass of the supermassive black hole ($M{\rm BH}$) was also added \citep{molla16}, following the classical expression: $\log{M_{\rm BH}} = \beta\,\log{(\sigma/220)}+\alpha$, with $\alpha \approx 8$ and $\beta \approx 4$ \citep{gerbhardt00}.  The central velocity dispersion, $\sigma = 50.0867\,R_{\rm D}^{0.9722}$, is obtained from correlations between the luminosity, the central velocity dispersion and the disc scale-length ($R_{\rm D}$) given in \citet{balcells07a}.  

Since Eq.~\ref{eq:mass_bul_r} gives the final bulge mass for each radial region, and from \citet{molla16} we known the mass in the halo for the same radial region, the collapse time-scale to form the bulge or disc can be calculated from Eq.~\ref{eq:tcoll}. Another important point is the radius that we should consider to calculate the bulge mass. By choosing this radius as being $R = 4R_{\rm e}$, Eq.~\ref{eq:mass_bul_r} results that 85\% of the bulge mass is inside this region. In the model $R_{\rm e}=0.536$\,kpc, so that $R = 4R_{\rm e}$ resulted in the value for the bulge radius $R_{\rm B}=2.1$ kpc, which is in agreement with the most recent determinations, as pointed above. This imposition can be done safely, since the contribution of the bulge mass for each subsequent ring $R > R_{\rm B}$ would be less than a 1\%. Therefore, to use other value for $R_{\rm B}$ does not change appreciably the results.

The resulting masses for each radial region in the disc, bulge and halo ($\Delta M_{\textrm{\scriptsize D,B,H}}(R)$, respectively) are displayed in Table~\ref{tab:tcoll}. It is also shown in this table, the total mass included in each cylindrical region above and below the corresponding radial region in which the gas will fall as defined in Eq.~\ref{eq:mtot}, as well as the collapse time-scales for each region, $\tau(R)$, obtained from Eq.~\ref{eq:tcoll}. 

\begin{table}
    \setlength{\tabcolsep}{3pt}
	\centering
	\caption{Characteristics of modelled bulge and inner disc radial regions.}
	\label{tab:tcoll}
	\begin{tabular}{cccccc} % four columns, alignment for each
		\hline
$R$ & $\Delta M_{\rm tot}$ & $\Delta M_{\rm D}$  & $\Delta M_{\rm B}$ & $\Delta M_{\rm H}$ & $\tau$ \\
(kpc) & ($10^9$~\Msun) & ($10^8$~\Msun) & ($10^9$~\Msun) & ($10^7$~\Msun) & (Gyr)\\
\hline
0.00 & 3.710 & 0.000 & 3.666 & 0.000 & 1.054 \\
0.25 & 5.002 & 0.309 & 4.971 & 0.048 & 1.425 \\
0.50 & 3.108 & 1.599 & 2.944 & 0.329 & 1.927 \\
0.75 & 2.309 & 3.438 & 1.956 & 0.880 & 2.370 \\
1.00 & 1.960 & 5.523 & 1.391 & 1.688 & 2.776 \\
1.25 & 1.831 & 7.688 & 1.035 & 2.742 & 3.142 \\
1.50 & 1.819 & 9.835 & 0.795 & 4.028 & 3.465 \\
1.75 & 1.870 & 11.887 & 0.626 & 5.536 & 3.750 \\
2.00 & 1.957 & 13.811 & 0.503 & 7.253 & 4.006 \\
3.00 & 7.610 & 71.103 & 0.000 & 49.930 & 4.846 \\
4.00 & 9.689 & 87.792 & 0.000 & 91.005 & 5.581 \\
5.00 & 10.800 & 94.013 & 0.000 & 139.880 & 6.458 \\
\hline
	\end{tabular}
\end{table}

\subsection{Stellar yields}
\label{yields}

The FRUITY\footnote{\url{http://fruity.oa-teramo.inaf.it/}} set \citep{cristallo11,cristallo15} is used as the stellar yields for low and intermediate-mass stars. \citet{cristallo11} give stellar yields for stars between 1 and 3\,\Msun\, and \citep{cristallo15} complemented the set for masses in the range from 4 to 6\,\Msun. Their models include the evolution of each star and, particularly, the AGB phases with thermal pulses and dredge-ups. In the FRUITY web page one can obtain the stellar masses and core and surface abundances in each step, as well as the net and total yields for most relevant isotopes and 10 metallicities. Their so-called total yields were used, which actually are the total ejected mass (new and already existing in the star when formed) of each isotope to include in our code. We have computed the secondary and primary components for $^{14}$N and $^{13}$C by assuming that the production during the third dredge-up may be considered as primary and the rest is secondary.

For massive stars, we employed in our models the stellar yields set from \citet{limongi18}. These authors give at their web page\footnote{\url{http://orfeo.iaps.inaf.it}} the stellar yields for four metallicities ([Fe/H]=0, -1, -2, -3) and for nine stellar masses, from 13 to 120\,\Msun. Besides, the yields set include stellar rotation for three different values of velocity: 0, 150 and 300\,km\,s$^{-1}$. \citet{prantzos18} empirically constrained metallicity-dependent weighted average for the yields, favouring faster rotation at lower metallicities. We followed their recommendations and adopted the distributions shown in their Fig.~4. The most important impact of the rotation in massive stars is the modification of the stellar yields for low metallicities, mainly the one for $^{14}$N, which appears as primary when [Fe/H]$\le -3$. We interpolated at the same 10 metallicities as FRUITY,  to obtain a set of ejected masses for the whole mass range from 1 to 120\,\Msun\, and 10 metallicities. We note that it is also needed an interpolation between 6\,\Msun, the most massive star in the FRUITY set, and 13\,\Msun, the less massive star in the \citet{limongi18} set. We follow again the scheme adopted by \citet{prantzos18}, assuming that for stars up to 10\,\Msun, stars evolve as AGBs, and their yields can be obtained by extrapolation of the low and intermediate-mass yields, weighted by the corresponding ejecta mass \citep[see equation 5 from][]{prantzos18}. A log-log interpolation is then performed between 10 and 13\,\Msun. Additionally, the stellar yield for $^{24}$Mg was multiplied by a factor of 2 to better reproduce the observational data. For the same reason, the Si and Ca stellar yields were multiplied by a factor 0.8 and 1.4, respectively. 

Finally, we added the stellar yields for Supernovae Ia (SNe-Ia). SNe-Ia releases are included by using the model revised by \citet{iwamoto99}. The $\alpha$-elements, where $\alpha$ stands for O, Ne, Mg, Si, S and Ca, appear in the ISM when massive stars explode as core-collapse supernovae (CC-SNe), after the first $\sim 3-30$\, Myr. Since Fe is ejected by both CC SNe and SNe-Ia events, the [$\alpha$/Fe] ratio is sensitive to the SN rate of each region within the Galaxy. Additionally, besides the yield for individual SNe, it is necessary to consider their rates of production. SNe-Ia progenitors were expected to reside in  binary stellar systems, which explode after the evolution of the most massive star. So that, there is a time delay before the Fe ejected by the SN Ia compared to the one created by CC SNe. As a consequence, for low [Fe/H], there is a plateau in the [$\alpha$/Fe], since Fe is mainly produced by CC SNe at that time. As the binary systems will evolving, the explosions of SNe-Ia start and there is an additional source of Fe, causing a {\it knee} in the  [$\alpha$/Fe] {\sl vs.} [Fe/H] diagram. We included this Delay Time Distribution (DTD), that describes how many type Ia SNe progenitors die after a delay time $\tau$ for a single stellar population of mass 1\,\Msun, using the DTD from  \citet{strolger20} obtained with parameters given in their figure 5, right panel. This DTD includes the scenario where two white dwarfs coalesce (double degenerate systems). For more details about this DTD see also the discussion presented in \citet{molla22}.

In this work we use {\sc starmatrix}\footnote{\url{https://github.com/xuanxu/starmatrix}} \citep{bazan22}, an Open Source Python implementing the Q-matrices technique to compute the new ejected elements by each single stellar population, later convolved with the SFR in the CEM. The adopted solar chemical composition is the one provided by \citet{asplund09}. 

\section{Model results}
\label{results}
\subsection{The infall rate}

Since in this model the infalling mass from the halo to the equatorial plane to form the bulge and disc components has different collapse time-scales, the inner regions of the bulge will be formed with a shorter collapse time-scale of $\tau \sim 1$~Gyr, meanwhile the outer regions will be formed with a longer time-scale $\tau\sim 3.5$~Gyr, characterizing an inside-out formation of the bulge. 

\begin{figure}
	% To include a figure from a file named example.*
	% Allowable file formats are eps or ps if compiling using latex
	% or pdf, png, jpg if compiling using pdflatex
	\includegraphics[width=8cm]{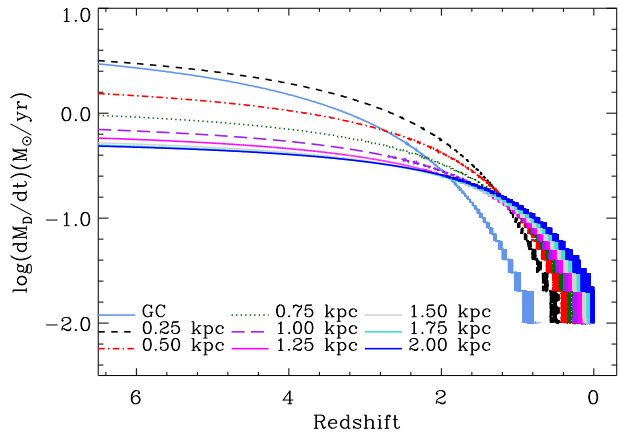}
    \caption{The infall rate in logarithmic scale as a function of the redshift $z$ for the modelled radial regions of the GB, as labelled.}
    \label{fig:infall}
\end{figure}

As a consequence of the variation of the collapse time-scale, the infalling rate is a function of the time and space. In Fig.~\ref{fig:infall} it is shown the infall rate as a function of the redshift for several radial regions located inside the bulge. In order to convert the time to redshift, we have considered the relation given by \citet{macdonald06} and the $\Lambda$CDM cosmology with the cosmological parameters $H_0=67.4$\,km \,s$^{-1}$\,Mpc$^{-1}$, $\Omega_M=0.315$ and $\Omega_\Lambda=0.685$ \citep{planck2018}. As can be seen in the figure, initially the infall rates at the central regions of the bulge are higher than in the outer ones. On the other hand, the outer regions have a prolonged infall compared with the inner regions. As we will show, the dependence of the infall rate with time and space has an important consequence in the metallicity distributions in the bulge. 

\subsection{The metallicity distribution functions}

\begin{figure*}
	\includegraphics[width=14cm]{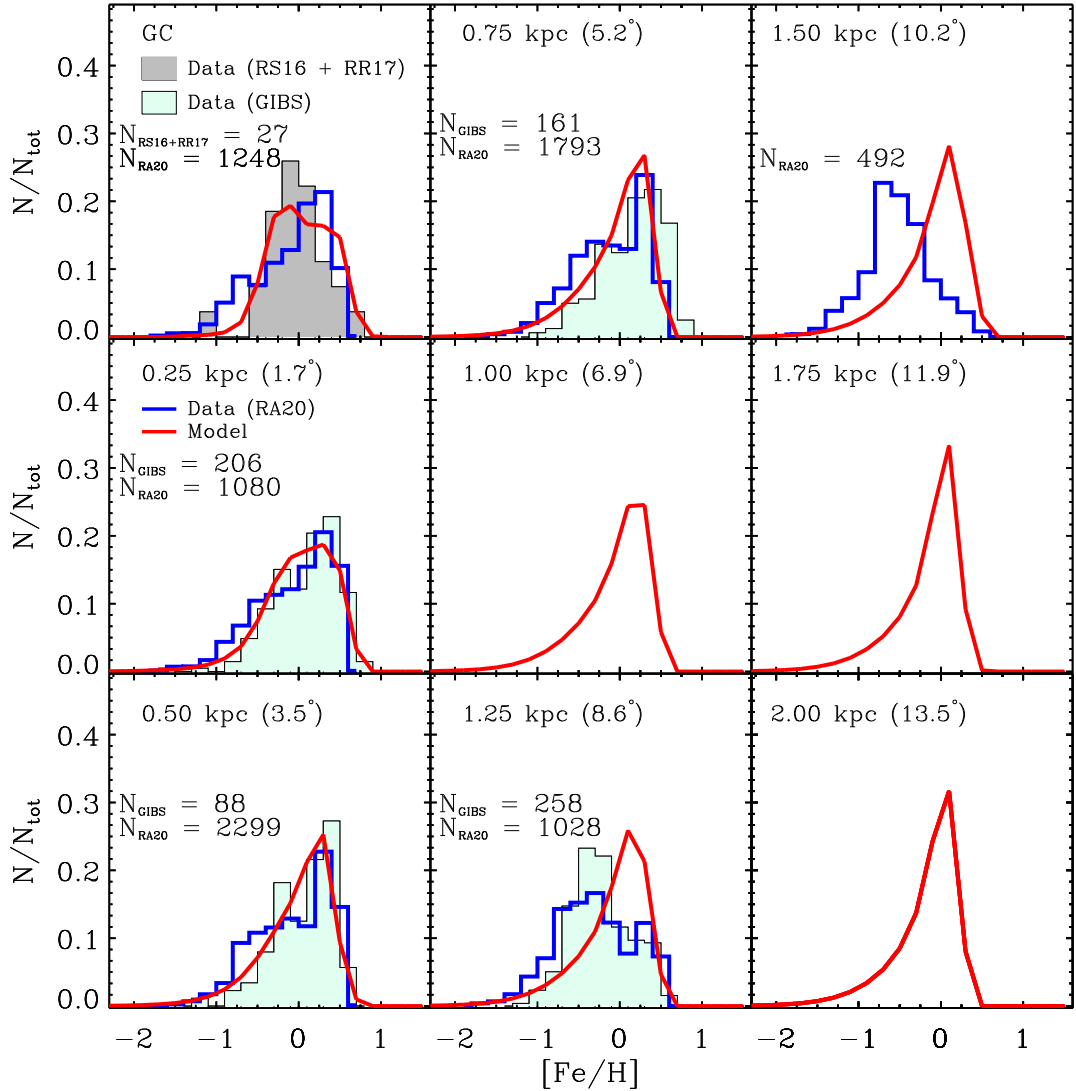}
    \caption{Model predictions for [Fe/H] histograms (unfilled red lines) for different regions in the GB (0.00, 0.25, 0.50, 0.75, 1.00, 1.25, 1.50, 1.75 and 2.00\,kpc), as labelled in each panel. The theoretical MDFs represent those from giant GB stars, being the stellar abundances computed taking into account the stellar life-times from Padova isochrones. Filled light-cyan histograms are the GIBS data from \citet{zoccali17}, except for the GC, where the data from \citet{ryde16} (RS16) and \citet{rich17} (RR17) are shown instead (filled gray histogram). The data from \citet{rojas-arriagada20} (RA20) are showed as blue unfilled histograms. The number of stars to build the histograms are shown at the middle left of each panel where the observational data are plotted. 
    }
    \label{fig:mdf}
\end{figure*}

The resulting GB metallicity distribution function (MDF) for each spherical region is shown in Fig.~\ref{fig:mdf}, represented by [Fe/H] histograms. 

The theoretical MDFs were constructed by computing the stellar abundances taking into account the model results plus the stellar mean life-times from Padova isochrones as used by \citet{molla09} and with the fits computed by \citet{raiteri96}, as well as the \citet{kroupa01} IMF. Furthermore, the resulting MDFs were convolved with a Gaussian to take into account a typical uncertainty of 0.15\,dex in the observational data. To validate the model, we show in the same Fig.~\ref{fig:mdf} the observational data from \citet{ryde16} (RS16) and \citet{rich17} (RR17) for the Galactic Centre (GC) panel, and from \citet{zoccali17} (GIBS) in the other panels corresponding to galactocentric distances 0.25, 0.50, 0.75 and 1.25\,kpc panels. The observations from \citet{rojas-arriagada20} (RA20) are also plotted as a blue line in the GC, 0.25, 0.5, 0.75, 1.25 and 1.5\,kpc panels. In order to compare with these data, we have converted the angles of the data to radial distances by adopting a Solar galactocentric distance of 8.3\,kpc \citep{gillessen09}. The corresponding angle in degrees for each radial region is displayed in each panel along with the corresponding distance $R$. 

One of the caveats to try to reproduce the MDFs from the observations is that their shapes are strongly affected by selection effects. For example, the selection function of the GIBS survey peaks approximately at 1\,kpc around the Galactic Centre and, consequently, the outer bulge fields should be compared with caution. 

Another important point to compare model and data is that GIBS observations of bulge fields very close to the Galactic plane may suffer from disc contamination. This may be the case of the field $\ell = +8$, $b=-2$, as discussed in \citet{zoccali17}. Therefore, it is  important to compare not only the peak positions of the MDFs, but also the metallicity range of the distributions. Last but not less important, at the current stage of the observational data, is not possible to have a truly 3D distribution of the MDF in the bulge, since accurate Gaia distances are not available for stars at these distances. Therefore, we only have a 2D projection of the bulge MDF in the $(\ell,b)$ plane and the comparison with the model may not be straightforward. We have to be aware of an additional effect about the circular symmetry adopted here to calculate the angular distances in ($\ell,b$) plane with respect to the GC: \citet{zoccali17} demonstrated (see their Fig.~9) that the spatial distribution of MR and MP stars are different. In particular, MR stars display a boxy shape in the sky, while MP ones display a more spherical distribution. In this context, the conversion angle-to-distance with circular symmetry around the GC, which we applied here, mix MR and MP stars in proportions that maybe are not the correct ones, since in reality they could be not a simple function of the angular distance from the GC, but also of the angle with respect to the midplane, especially for angular distances from the GC higher than $3\degr$. Below $3\degr$ from the GC, the circular symmetry around the GC applied here is not strongly affected by this effect, since both metal-rich and metal-poor stars are present.

The model predicts a wider MDF for the GC and the 0.25\,kpc regions, while it starts to become increasingly narrower for the outer regions, a tendency also corroborated by the observational data. 
\begin{figure*}
	\includegraphics[width=14cm]{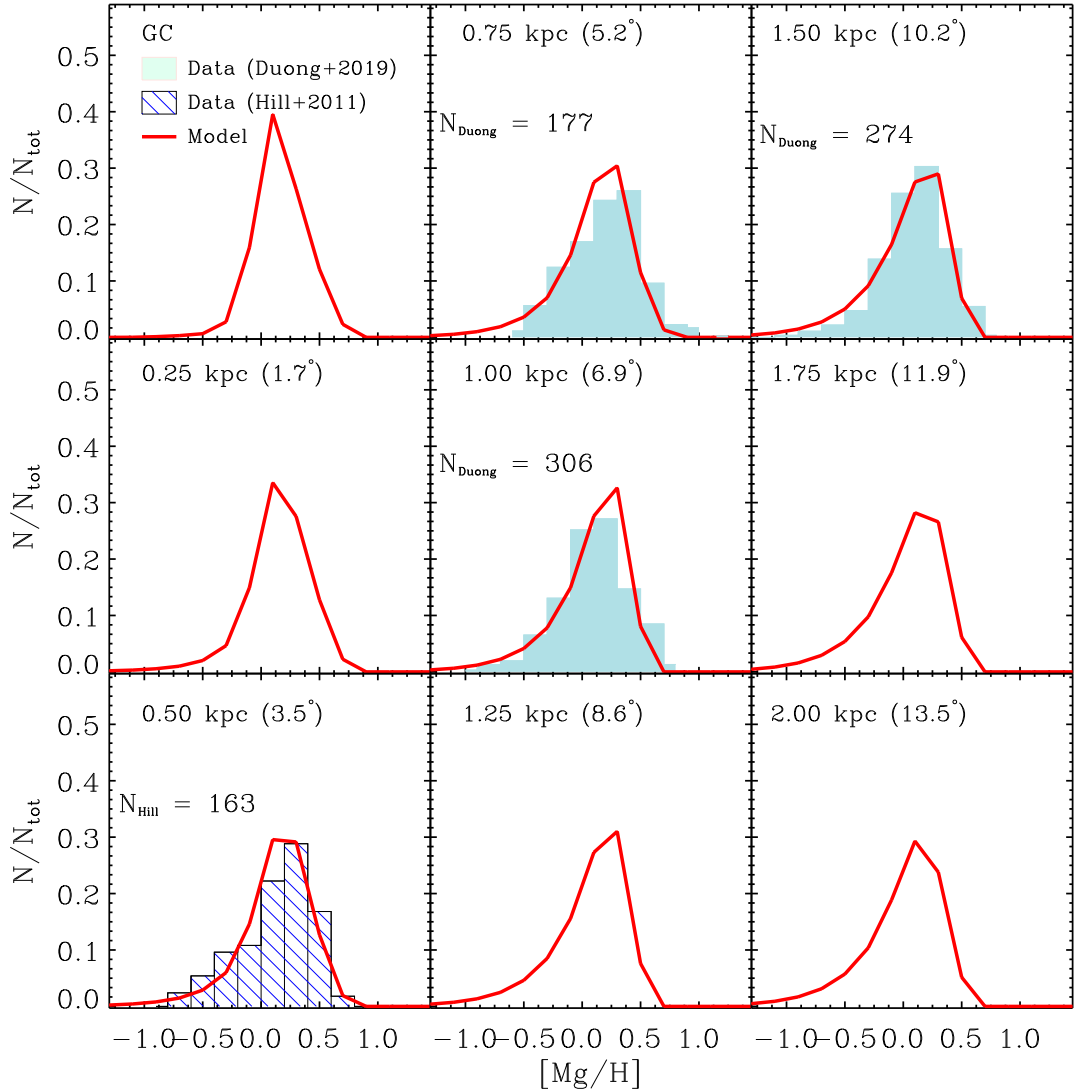}
    \caption{[Mg/H] histograms for different regions in the GB (0.00, 0.25, 0.50, 0.75, 1.00, 1.25, 1.50, 1.75 and 2.00\,kpc), as labelled in each panel. The theoretical histograms represent those from giant GB stars, being the stellar abundances computed taking into account the stellar life-times from Padova isochrones. The data from \citet{duong19} are shown at the 0.75, 1.00 and 1.50\,kpc panels. At the 1.25\,kpc panel, it is also shown the observational data from \citet{hill11} to illustrate the comparison with the model.}
    \label{fig:mghhist}
\end{figure*}
The model predicts that the number of stars composing the MP population is higher in the central fields than in the outer ones, which is confirmed by the data \citep{zoccali17,zoccali18}. 

Overall, the model is able to reproduce quite well the observational data, specially at the inner regions, where the selection function of the observational data from the GIBS survey peaks. As noted by many other observational studies and \citet{zoccali17}, the data show double peaked histograms in some fields, (as in $R=1.25$\,kpc panel), suggesting a GB composed by two populations with different mean metallicities. In the case of \citet{rojas-arriagada20}, the data show even a trimodal MDF (not seen here because of the adopted bin size and the composition of the selected fields). Our model shows a slight bimodality at the GC panel, with a higher proportion of MP stars (55\% {\sl vs.} 45\%) as observed, but is not able to reproduce this feature at the outer regions. Probably, it is necessary another scenario for those regions, as e.g. a radial inflow of gas coming from the inner disc due to the secularly evolved disc into a boxy/peanut X-shape bar, as noted by many studies \citep{ness12,bensby17,rojas-arriagada17,zoccali17}. 
In the outermost spherical shell (1.5\,kpc, 10.2$\degr$) where data is available, an overlap between the disc and bulge population must exist, and,  however, the two component populations are not longer observed. We note that this sample should closely correspond to the last latitudinal bin ($|b|>10$) in Fig.~9 of \citet{rojas-arriagada20}. As they explain, this sample might be dominated by thick disc stars and, therefore, both the discrepancy with the CEM and the simple shape of the observed MDF may be due to this effect. This way, differences  found for this latitudinal bin compared with the other panels are not very important for the analyses performed here.

Nonetheless, it is evident in this Fig.~\ref{fig:mdf} that there are differences in the MDFs for each region. In the model, the fraction of the metal poor population is somewhat higher than the metal rich one in the innermost region (GC), which is in agreement with the data from RS16 and RR17. Despite the low number of stars in these samples in RS16 and RR17 the observations were performed in the $K$ band ($\sim 2.2\, \mu$m) with VLT and Keck telescopes, respectively, which allows to go deeper into the high extinction fields of the bulge, and thus represent more accurately the GC population than the Apogee observations, which were performed at $H$ band ($\sim 1.65\, \mu$m) by using telescopes of 2.5\,m class. On the other hand, the MDF of the Apogee observations peaks at higher metallicities than RS16 and RR17 data and also than the model prediction for the GC region. 

Moving further out to the other spherical shells, in panels $1.7\degr$, $3.5\degr$ and $5.2\degr$ the observations show that the fraction of metal-poor stars increases as the angular distance to the GC increases. The model is not able to reproduce this trend in much detail. In spite of that, the model predicts that the positions of the metal-rich peak do not change considerably, which is in agreement with the observations. In all panels where the observational data are available, the lower and upper limits of the MDF are well reproduced by the model, indicating that the predicted time-scales are probably correct.

In Fig.~\ref{fig:mghhist} we now show the magnesium histograms in the notation [Mg/H]. The models are compared with the stellar data from \citet{hill11} and \citet{duong19}. The last one provides alpha-element abundances along the minor bulge axis at latitudes $b = -5{\degr}, \, -7.5{\degr}  \, {\rm and\, } -10{\degr}$. We again converted this fields to galactocentric distances and considered them equivalent to the regions 0.75, 1.00 and 1.50\,kpc, respectively. On the other hand, in \citet{hill11} the  number of stars is reduced and there is no spatial distribution for this stars in the bulge. Their sample corresponds to red clump stars observed in the Baade's Window, which is located in the minor axis at $b\sim-4\degr$. The closest simulated region in our CEM is 0.5\,kpc (3.5$\degr$), so that their data are represented in this panel. In general, we found a good correspondence between data and model in this comparison. In the panel 0.75\,kpc (5.2$\degr$), the data indicates a slightly higher number of metal-rich stars than our model predicts. We should highlight that the observational data from \citet{duong19} are compatible with regions where the time-scales are longer than in others, which would imply that probably the observed stars are located in the outer parts of the bulge. However, in order to have a better comparison with the models, we need accurate distances for the bulge stars to properly have a 3D distribution of these stars. 

\begin{figure}
	\includegraphics[width=8cm]{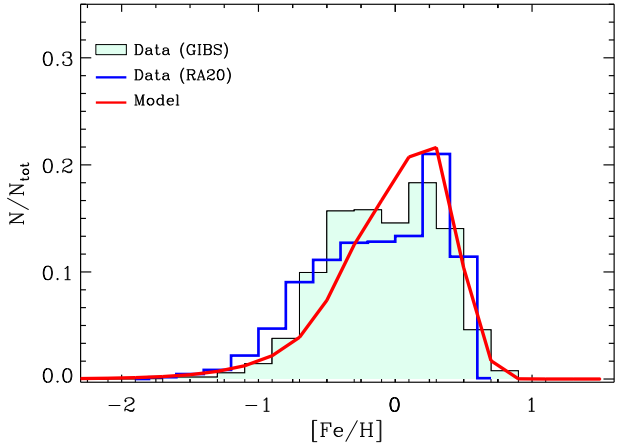}
   \caption{The global bulge MDF computed by summing the stars in all the simulated regions within the bulge.}
    \label{fig:mdf_all}
\end{figure}

In order to compare the effects of the inside-out bulge formation in the MDF we computed in Fig. \ref{fig:mdf_all} a single MDF by including the stars in all the simulated bulge regions. For this plot the MDF was built by summing the stars in all the simulated bulge regions. We also added all the stars for each dataset separately and plotted in the figure for a comparison with the model. As can be seeing, the distributions are asymmetrical for both data and model. The distributions peak at [Fe/H] $\sim$0.25~dex and drop rapidly for higher metallicities. On the other hand, the lower metallicity tail is smoother. The data show an slightly excess of stars with [Fe/H] < 0.0~dex compared with the model. However, the lower and upper limits of the distributions are very well reproduced by the chemical evolution model. From this comparison we can interpret the asymmetrical bulge MDF as the result of combining stellar populations that where formed in distinct regions within the bulge formed with different collapse time-scales.  

\subsection{Radial distributions of abundances}

Another important question that we should address is whether or not the radial gradient of abundances observed in the disc is constant as approaching the galactic centre. This is an open problem in the literature, since until now the uncertainties in the distances do not allowed to obtain firm conclusions. There are evidences that the mean metallicity of the central regions of the Galaxy is not compatible with the extrapolation of the disc radial gradient towards the center \citep[see eg.][]{schultheis19}. In a recent paper from \citet{queiroz21}, the combination of APOGEE DR16 stars plus Gaia EDR3 parallaxes shows a spatial dependency of the metallicity in their Fig.~7, with a metal poor ($\alpha$-rich) component that seems to dominate the more central region. So that, recent observational results about the chemical abundances in the Galactic bulge indicate that the radial gradient in this region is not an extrapolation of the disc radial gradient. 
\begin{figure}
	% To include a figure from a file named example.*
	% Allowable file formats are eps or ps if compiling using latex
	% or pdf, png, jpg if compiling using pdflatex
	\includegraphics[width=8cm]{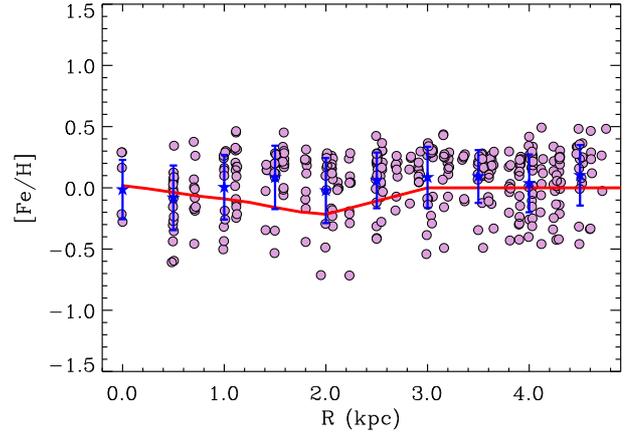}
    \caption{The metallicity, represented by [Fe/H], as a function of the Galactocentric distance. The model predictions are represented as a continuous red line. The observational data for the bulge and inner disc are from \citet{garcia-perez18}, displayed as filled magenta circles. Blue stars with error bars represent the bins of 1\,kpc wide and the standard deviation of the observational data, respectively. The bulge region in the model corresponds to $R < 2$\,kpc.
    \label{fig:fe_grad}
    }
\end{figure}
In this work, the disc component is modelled as described in \citet{molla22}, by using the infall prescription described in \citet{molla16}, and by including the new stellar yields and DTD function given in Section~\ref{yields}. The bulge is modelled as reported in section ~\ref{model} from this paper.
In Fig.~\ref{fig:fe_grad}, we compare the results for the [Fe/H] radial distribution in the bulge and inner disc with observational data from \citet{garcia-perez18}. In their analysis, the bulge parts that are further from the Sun beyond the GC ($R > 0$ in their notation) are susceptible to selection effects and biases and we opted to exclude these stars in Fig.~\ref{fig:fe_grad}. The model predicts that there is a radial [Fe/H] gradient of -0.12\,dex\,kpc$^{-1}$ within the bulge region (equivalent to -0.02 dex\,degree$^{-1}$). This radial gradient is very difficult to infer from the observations, given the spread in the [Fe/H] {\sl vs.} $R$ plane, which may be attributed to both uncertainties in metallicities ($\sim$ 0.12\,dex) and in distances ($\sim 20\%$). At the bulge distance, $20\%$ uncertainty in distances implies typical errors of $\pm 1.6$\,kpc, which contributes to the spread seen in the figure. The radial gradient found in Fig.~\ref{fig:fe_grad} is also supported by the observational data from \citet{gonzalez15}, where the authors find that the bulge has a gradient that is due to the different mix of the MR and MP populations and that it is mostly vertical rather than radial. The radial gradient found in this work is also much shallower than the one in the vertical direction of $\sim -0.6$\,dex\,kpc$^{-1}$ from \citet{zoccali08}. Furthermore, there exists other reported measurements of a metallicity gradient in the bulge minor axis of the order of -0.3 to -0.7\,dex\,kpc$^{-1}$ \citep[e.g.][]{minniti95,zoccali08,gonzalez13,rojas-arrigada14}. In particular, the gradient found in this work is in agreement with the observational result of -0.04 dex/degree (equivalent to -0.3 dex/kpc) from \citet{gonzalez13}. Nonetheless, the observational results reported in the literature correspond to a vertical gradient, not necessarily equivalent to a radial one.

\begin{figure}
	% To include a figure from a file named example.*
	% Allowable file formats are eps or ps if compiling using latex
	% or pdf, png, jpg if compiling using pdflatex
	\includegraphics[width=8cm]{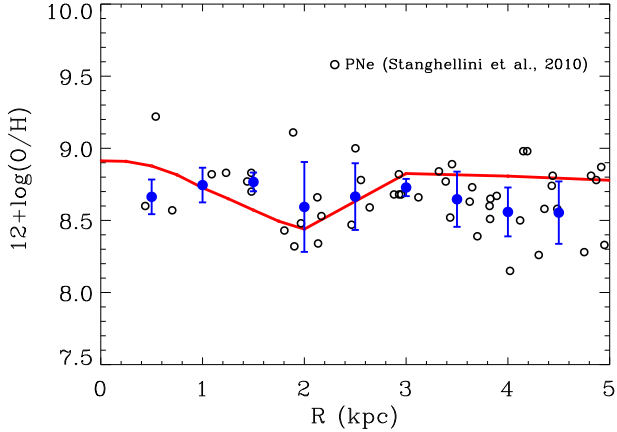}
    \caption{ The Oxygen abundance, in the notation $\log({\rm O/H})+12$, radial distribution in the bulge and inner disc.  The model predictions are represented as a continuous red line. The observational data from \citet{stanghellini10} for planetary nebulae (PNe) are displayed as black unfilled circles. Blue stars with error bars represent bins of 1\,kpc wide and the standard deviation of the observational data, respectively. The bulge region in the model corresponds to $R < 2$\,kpc. 
    \label{fig:oh_grad}
    }
\end{figure}

Oxygen abundances for bulge stars from these surveys are not available in the literature. This way, in order to compare the $\alpha$-abundance gradient obtained with the model, represented in Fig.~\ref{fig:oh_grad} with a red line, we have used the observational data for planetary nebulae (PNe) from \citet{stanghellini10}. We see a good correspondence between model and data in this figure, being that the observational data seems to have a flat or slightly negative gradient for  $3 < R < 4$\,kpc, and a positive gradient for $2 < R < 3$\,kpc. Then, there is an increase in the abundances by $\approx$ 0.3\,dex and the gradient becomes flat or slightly negative 
for $R < 2$ kpc. Although the low number of the PNe for $R < 2$\,kpc prevents to draw firm conclusions about the gradient at this region and more data should be observed. 
To these facts, it is necessary to add that PNe abundances represent the enrichment state not at the present time but at least $\sim 2$\,Gyr ago \citep{maciel13}.

\subsection{The alpha-element abundances over the iron abundance}

Higher Higher $[\alpha/{\rm Fe}]$ ratios at low [Fe/H] indicate a fast enrichment by massive stars,  which explode as type II  supernovae, that is a rapid formation of the stellar population. On the other hand, at higher metallicities ([Fe/H]$ > -0.5$\,dex) the type Ia supernovae have already enriched the interstellar medium, indicating a slower stellar formation process. Therefore, the star formation histories of the different regions  have a reflect  on the relative abundances $[\alpha/{\rm Fe}]$ relative abundances. 

\begin{figure}
	\includegraphics[width=8cm]{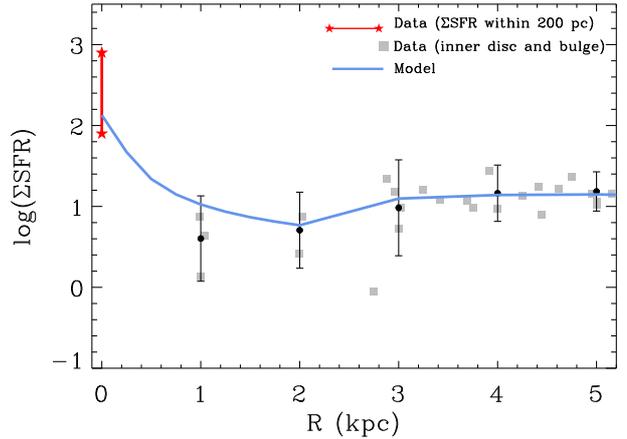}
   \caption{Comparison between the predictions of our model for the radial profile of the star formation rate surface density ($\Sigma$SFR) and the observational data for the bulge and inner disc. See the text for details and references.}
    \label{fig:sigma_sfr}
\end{figure}

The GC region presents an intense star formation process and, consequently, formed a higher fraction of massive stars than in the other bulge regions at the early times. In \citet{grieco15} a CEM is built to model the GC and the conclusion was that this region formed in a short time-scale and fast episode of gas infall. This model was subsequently used in \citet{ryde16}, where it was necessary to invoke a top-heavy IMF to reproduce the observational data. In Fig.~\ref{fig:sigma_sfr}, we show the radial profile of the SFR surface density ($\Sigma_{SFR}$ in units of \Msun\,Gyr$^{-1}$\,pc$^{-2}$) compared with data for the disc and inner bulge as compiled in \citet{molla15}. These data were a compilation independently derived from various methods and authors \citep{lacey83, williams97,peek09,urquhart14}. The figure also shows the GC data that are in the range of  $\log(\Sigma_{SFR}) \sim$ 1.9 to 2.9\,dex in the first 200\,pc from the GC, as given in \citet{henshaw22}. In this figure, black circles with error bars correspond to the mean and standard deviation of the inner disc and bulge data in bins of 1\,kpc wide. The agreement between model and data is excellent, given the observational dispersion, although we would need more data in the bulge region to confirm our predictions.

\begin{figure*}
	% To include a figure from a file named example.*
	% Allowable file formats are eps or ps if compiling using latex
	% or pdf, png, jpg if compiling using pdflatex
	\includegraphics[width=14cm]{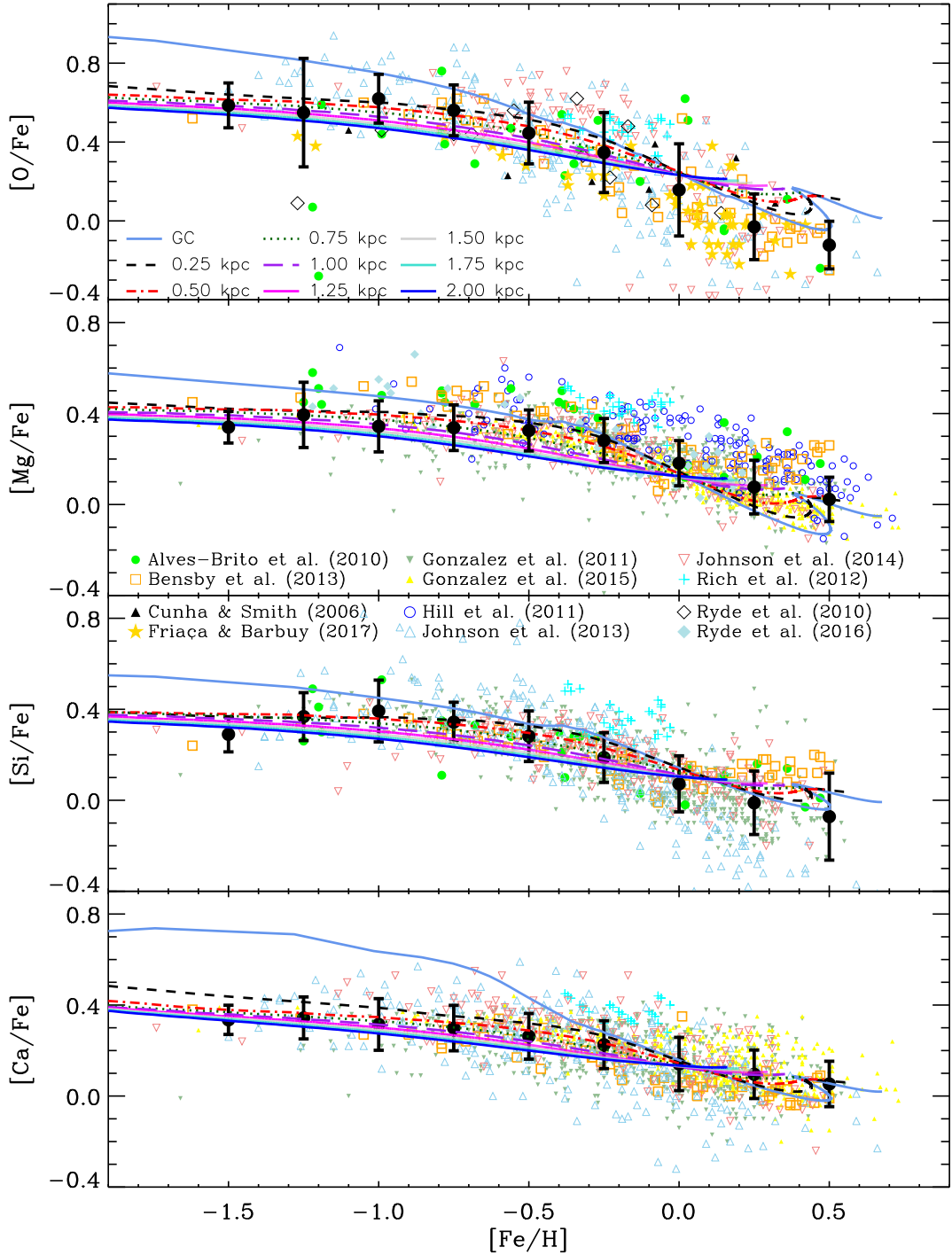}
	%ohmgcasifehvs_feh_12.010_classic_CRI_LIM_2018_mix_kroupa2001_mup_100_snia_strolger-fit5_dtd_corr_1.0_iwa1998_bin_frac_0.15_gamma_sec_2.0corr_mg_2.0_corr_si_0.8_corr_ca_1.4_corr_fe_1.0_epskbul_0.009_epsksol_0.009_epsh_0.6_epsa_0.2_eps1a_2.5e-8_opch2_1
    \caption{Comparison between the predictions of our model at different radial regions and the literature data for $[\alpha/\mbox{Fe}]$  where $\alpha$ stands for O, Mg, Si, and Ca,  {\sl vs.} [Fe/H]. The line codes of the models are as labeled in the first panel (from top to bottom). The observational data for the bulge are labelled in the other panels, and correspond to filled green circles from \citet{alves-brito10}, open squares from \citet{bensby13}, open upside black triangles from \citet{cunha06}, filled gold stars from \citet{friaca17}, filled blue upside-down triangles from \citet{gonzalez11}, filled upside yellow triangles from \citet{gonzalez15}, cyan crosses from \citet{rich12}, and filled and open diamonds from \citet{ryde10,ryde16}. Black dots with error bars are, respectively, the mean and standard deviation of the observational data in bins of 0.25\,dex wide.  
}
    \label{fig:alpha_fe}
\end{figure*}
In Fig.~\ref{fig:alpha_fe}, we shown the relative abundances $[\alpha/{\rm Fe}]$ {\sl vs.} the iron abundance $[{\rm Fe}/{\rm H}]$, where $\alpha$ stands for the $\alpha$-elements: O, Mg, Si and Ca. Overall, our model is able again to reproduce the observational data regarding the $[{\rm O}/{\rm Fe}]$, $[{\rm Si}/{\rm Fe}]$  and $[{\rm Ca}/{\rm Fe}]$ abundance ratios.  In our case, as explained above, we adopted the yields for massive stars from \citet{limongi18} and the stellar yield for $^{24}$Mg was multiplied by a factor of 2 to better reproduce the observational data. For the same reason, the Si and Ca stellar yields were multiplied by a factor 0.8 and 1.4, respectively. Even though, the $[{\rm Ca}/{\rm Fe}]$ relative  abundance data from \citet{bensby13} --open orange squares-- and \citet{rich12} --cyan crosses -- are slightly higher than the results of the model, except for the GC. Interestingly, the model predicts different $[\alpha/{\rm Fe}]$ ratios for each spherical region, the ratio being higher at the innermost regions (0.25 and 0.75\,kpc) and lower at the outermost regions, especially at low $[{\rm Fe}/{\rm H}]$ abundances, as expected for regions with different star formation histories. In this region of low metallicity the differences in $[\alpha/{\rm Fe}]$ ratios of the models are less important, probably because the SFR is already low in all radial regions.

The results of Fig.~\ref{fig:alpha_fe} suggest that the spread in $[\alpha/{\rm Fe}]$ ratios, as shown by the observational data at lower metallicities, may be the result of mixing stars from different radial locations within the GB. \citet{rojas-arriagada19} did the first estimation of the dispersion in the $[\alpha/{\rm Fe}]$ ratio in the literature. They computed $\sigma[{\rm Mg}/{\rm Fe}] = 0.031$\,dex for the dispersion of the metal-poor bulge stars in the metallicity range -1.0 < [Fe/H] < 0.1. To compare with their results, we calculated the standard deviations for each element, computed as the mean of the standard deviations for each value of [Fe/H]  in the range -1.0 < [Fe/H] < 0.1 \citep[the same metallicity range as in][]{rojas-arriagada19}. Table~\ref{tab:mean_std} shows the results for each element considered in the CEM. The first line in this table corresponds to the dispersion computed from all the nine radial regions and the second one those computed excluding the GC region. O and Mg show higher dispersions, while for Si and Ca the dispersions are lower and comparable. As can be seen, our estimation for $\sigma[{\rm Mg}/{\rm Fe}]$ is higher than the value of 0.031 dex from \citet{rojas-arriagada19}. As a test, we excluded the GC region from the computed dispersion and the results are lower than in the previous case for all the elements (All but GC in the table). It is important to note that the regions we are simulating, until $R_{B} = 2$ kpc, not necessarily cover the observations since the selected likely bulge stars criteria in \citet{rojas-arriagada19} are based on a radial and distance to the Galactic plane cut: $R < 3.5$\,kpc and $|z| < 0.5$\,kpc.
\begin{table}
	\centering
	\caption{Mean dispersion in the abundances in the metal-poor bulge.}
	\label{tab:mean_std}
	\begin{tabular}{lcccc} % four columns, alignment for each
		\hline
		    & $\sigma[{\rm O}/{\rm Fe}]$ & $\sigma[{\rm Mg}/{\rm Fe}]$ & $\sigma[{\rm Si}/{\rm Fe}]$ & $\sigma[{\rm Ca}/{\rm Fe}]$ \\
		    \hline
	    All regions & 0.091 & 0.079 & 0.052 & 0.051 \\
	    All but GC & 0.079 & 0.067 & 0.045 & 0.044 \\
		\hline
	\end{tabular}
\end{table}

\begin{figure*}
	% To include a figure from a file named example.*
	% Allowable file formats are eps or ps if compiling using latex
	% or pdf, png, jpg if compiling using pdflatex
	\includegraphics[width=14cm]{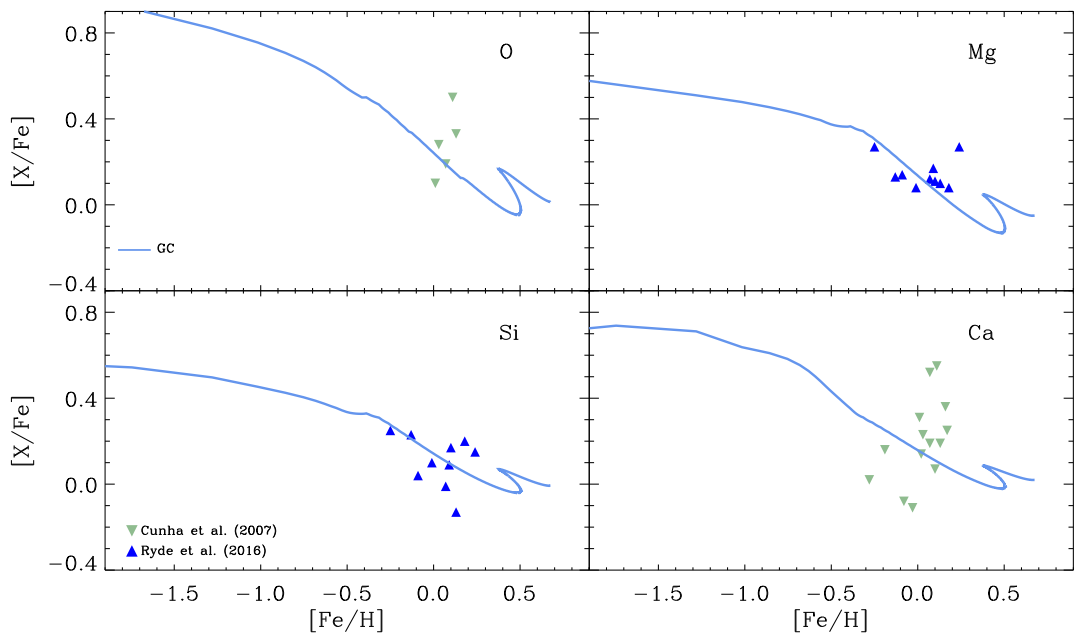}
	%ohmgcasifehvs_feh_12.100_classic_CRI_LIM_2018_mix_kroupa2001_mup_100_snia_strolger-fit5_dtd_corr_1.0_iwa1998_bin_frac_0.15_gamma_sec_2.0corr_mg_2.0_corr_si_0.8_corr_ca_1.4_corr_fe_1.0_epskbul_0.009_epsksol_0.009_epsh_0.6_epsa_0.2_eps1a_2.5e-8_opch2_1
    \caption{Comparison between the predictions of our model for the chemical evolution of the Galactic Centre and the literature data for [X/Fe], where X stands for O, Mg, Si, and Ca,  {\it vs.} [Fe/H]. The observational data for the bulge are labelled in the bottom left panel filled upside triangles from \citet{ryde16} and upside-down triangles from  \citet{cunha07}.}
    \label{fig:alpha_fe_centre}
\end{figure*}

There are previous attempts in the literature investigating the [$\alpha$/Fe] across the GB regions. For instance, \citet{duong19} investigated the [$\alpha$/Fe] trend with respect to the Galactic latitude and the results were not conclusive. Some differences were found at the high-metallicities regime, however the number of stars at these metallicities is small, and this bias can lead to wrong conclusions. Since the analysis is restricted to the bulge minor axis and is not fully 3D (only projections in the ($\ell$,$b$) plane are available), the small variation in [$\alpha$/Fe] found in the GCE model could not have been detected in their survey. Recently, \citet{griffith21}, using the APOGEE data, have explored the trend in the [$\alpha$/Mg] {\it vs.} [Mg/H] plane across three bulge regions, subdivided in groups according to the distance from the GC ($R$): $R < 1~{\rm kpc}$, $1~{\rm kpc } \le R < 2~{\rm kpc}$, $2~{\rm kpc } \le R < 3~{\rm kpc}$. They computed the median of the [$\alpha$/Mg] ratios in the three groups and find no differences within the uncertainties. Since their data are not publicly available, we cannot compare our models with their results. \citet{zasowski19} also analysing APOGEE data for [Mg/Fe] {\it vs.} [Fe/H] for different distances from the GC ($R$) in bins centered at $R=$ 0.8, 1.6, 2.4, 3.2 and 4.0 kpc. They investigated the {\it knee} (downturn) position and find a small dependence to lower metallicities at larger $R$. They concluded that this small difference might be due to significant radial mixing since the GB and disc formation that would erase any initial trends due to gradients in the SFR. 
To confirm the possible gradient of $[\alpha/{\rm Fe}]$ in the GB predicted by our models, we would need more data with higher precision, mainly for lower metallicities and galactocentric distances. However, we should also stress that, as visually shown, models are in agreement with the data dispersion presented at lower metallicities, [Fe/H] < 0, in Fig.~\ref{fig:alpha_fe}.

In Fig.~\ref{fig:alpha_fe_centre}, the [X/Fe] abundances as a function of [Fe/H] for our model, only for the GC, are plotted, where X stands for O, Mg, Si and Ca. For the elements specifically showed in this figure, the model with a normal IMF is able to reproduce the observational data, which is again an indicative that the collapse time-scale adopted for this region is in agreement with observations, although the small amount of data prevents us from drawing firm conclusions. We should highlight that in the CEM presented here it is not necessary to adopt a different IMF from the rest of the bulge to reproduce the GC data. We confirm that it is necessary a shorter time-scale in this region than in the other of the bulge.

\begin{figure}
	\includegraphics[width=8cm]{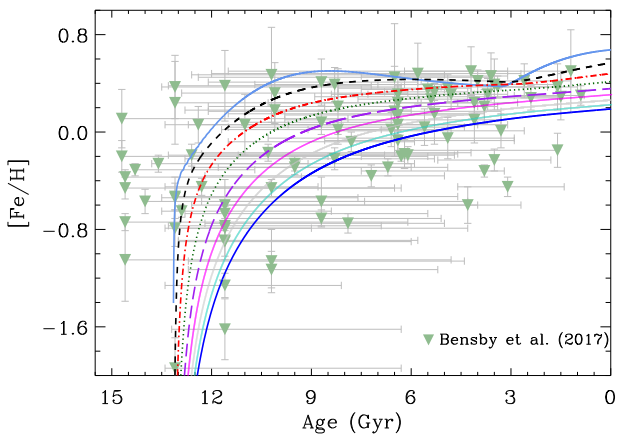}
   \caption{[Fe/H] as a function of the age in Gyr. The present time corresponds to 0\,Gyr. Models lines and colours are as in the previous figures. The data are from \citet{bensby17}.}
    \label{fig:feh_time}
\end{figure}

\subsection{Age evolution}

Fig. \ref{fig:feh_time} displays the plot [Fe/H] versus age, where it is also shown a comparison with the stellar data from \citet{bensby17}. We converted time in age in the model considering that the present time of 13.2 Gyr corresponds to the present time in which the chemical evolution is computed. Therefore, in this figure and in the following ones the age of 0\,Gyr is equivalent to the present time. The model predicts an age-metallicity relation in the sense that stars with low [Fe/H] are old, while stars with solar abundances or above span a wide range of ages. For example, taking the metallicity [Fe/H] $\sim$ -0.8 dex, it is expected the existence of stars with ages in the interval from $\sim$13 -- 11 Gyr. Meanwhile, for [Fe/H] $\sim$0.2 dex, a very wide range of ages in the interval 11.5 -- 0.0 Gyr is expected. This is somewhat the observational data show, nonetheless the uncertainties in the ages are very high. Therefore, in a model where the bulge is formed by different time-scales and not in a single and fast collapse, as commonly adopted in the previous chemical evolution models, there is an age-metallicity dependence which is similar to the observed data, considering the uncertainties. \citet{bernard18} also find a similar age-metallicity relation to \citet{bensby17}. They find that the fraction of stars younger than 5 -- 8 Gyr has a dependence with metallicity, according to their figure 11. For [Fe/H] < 0.0 the fraction of stars younger than 5 -- 8 Gyr is lower than stars with [Fe/H] > 0.0. An equivalent result is also found by \citet{hasselquist20}, strengthening these observational results.

\begin{figure}
	\includegraphics[width=8cm]{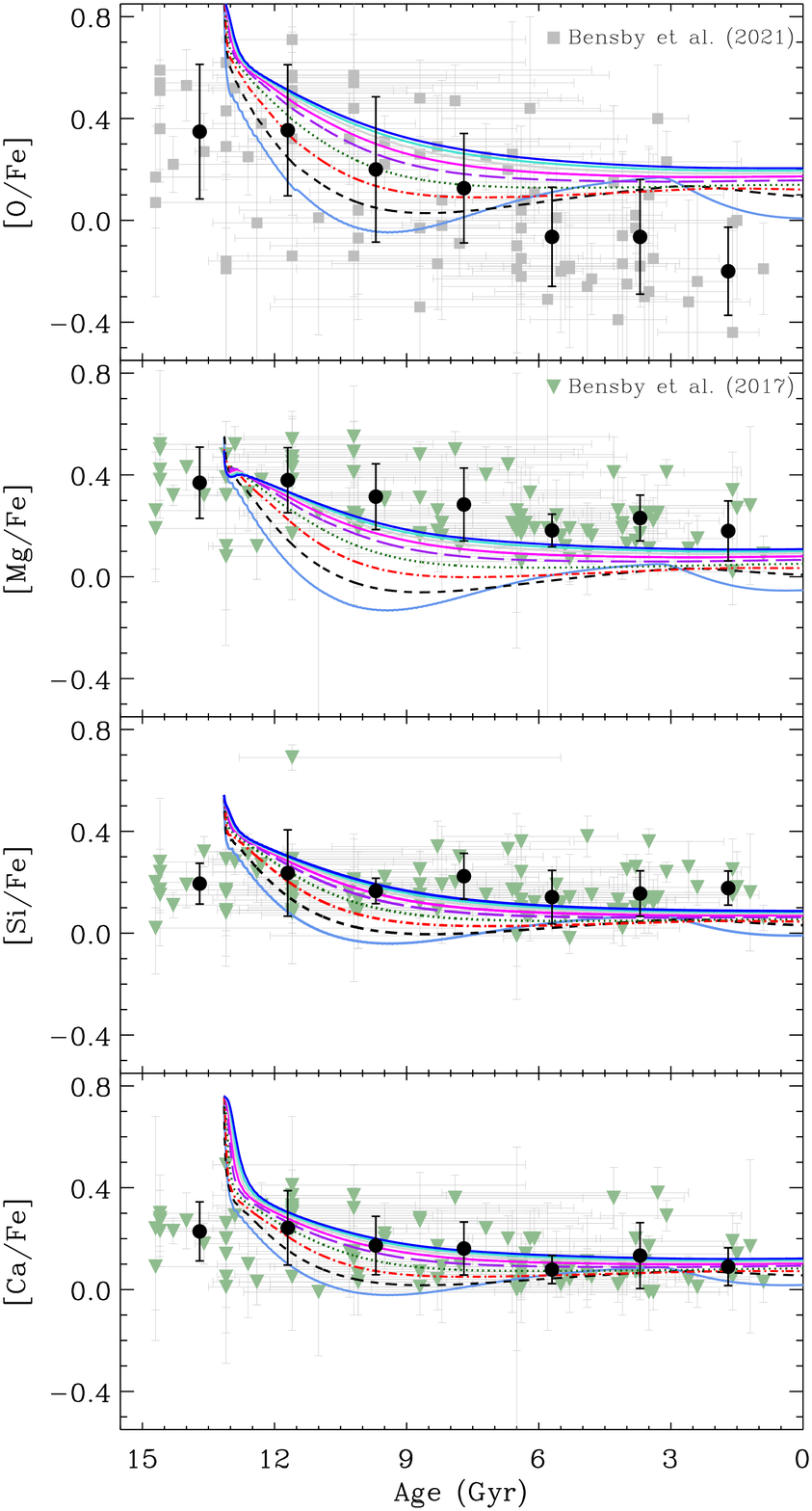}
   \caption{[$\alpha$/Fe] ratios as a function of age in Gyr. In the [O/Fe] plot the data from \citet{bensby21} are shown as grey squares. In the remaining plots the data from \citet{bensby17} are shown as upside-down triangles with error bars. Black circles with error bars are the mean and the standard deviations for bins of 2 Gyr wide for these data. The models are plotted with lines and colours as in the previous figures.} 
    \label{fig:ohmgsicafeh_time}
\end{figure}

In Fig. \ref{fig:ohmgsicafeh_time} we show the [$\alpha$/Fe] {\it vs.} age for O, Mg, Si and Ca. The models are compared with the data from \citet{bensby17}, except for [O/Fe], where the observations are from \citet{bensby21}. The models predict that at the beginning of the bulge formation (ages $\sim$12 Gyr) the [$\alpha$/Fe]  were higher than recently, since the main contributors to the Fe (the explosions of SNe-Ia) start later. At the same age, the [$\alpha$/Fe] for the central regions of the bulge are lower than the outer ones, because the first ones start the formation with higher infall rates. In this figure a good correspondence for Si and Ca predictions and the observational data appears. However, the model predictions for [Mg/Fe] are lower than the data. The predictions for [O/Fe] agree with the data for ages higher than 6\,Gyr and are above the data for ages lower than this value. We should note, however, that the data for [O/Fe] have a higher dispersion than the remaining ratios. The dependence of [$\alpha$/Fe] with age is not strong in the models, however older ages seems to have high [$\alpha$/Fe]. Looking for these trends in the data, we have binned the data in bins of 2\,Gyr wide and computed the mean and the standard deviation for each bin and the results are shown in the plots with black circles and error bars, respectively. We see that for ages older than $\sim$6\,Gyr [$\alpha$/Fe] are higher than for ages younger than this limit. The dispersion of the data is also higher for ages older than $\sim$6\,Gyr. Although the $\alpha$-ratios are below the data, these results are reproduced by the models.

\begin{figure}
	\includegraphics[width=8cm]{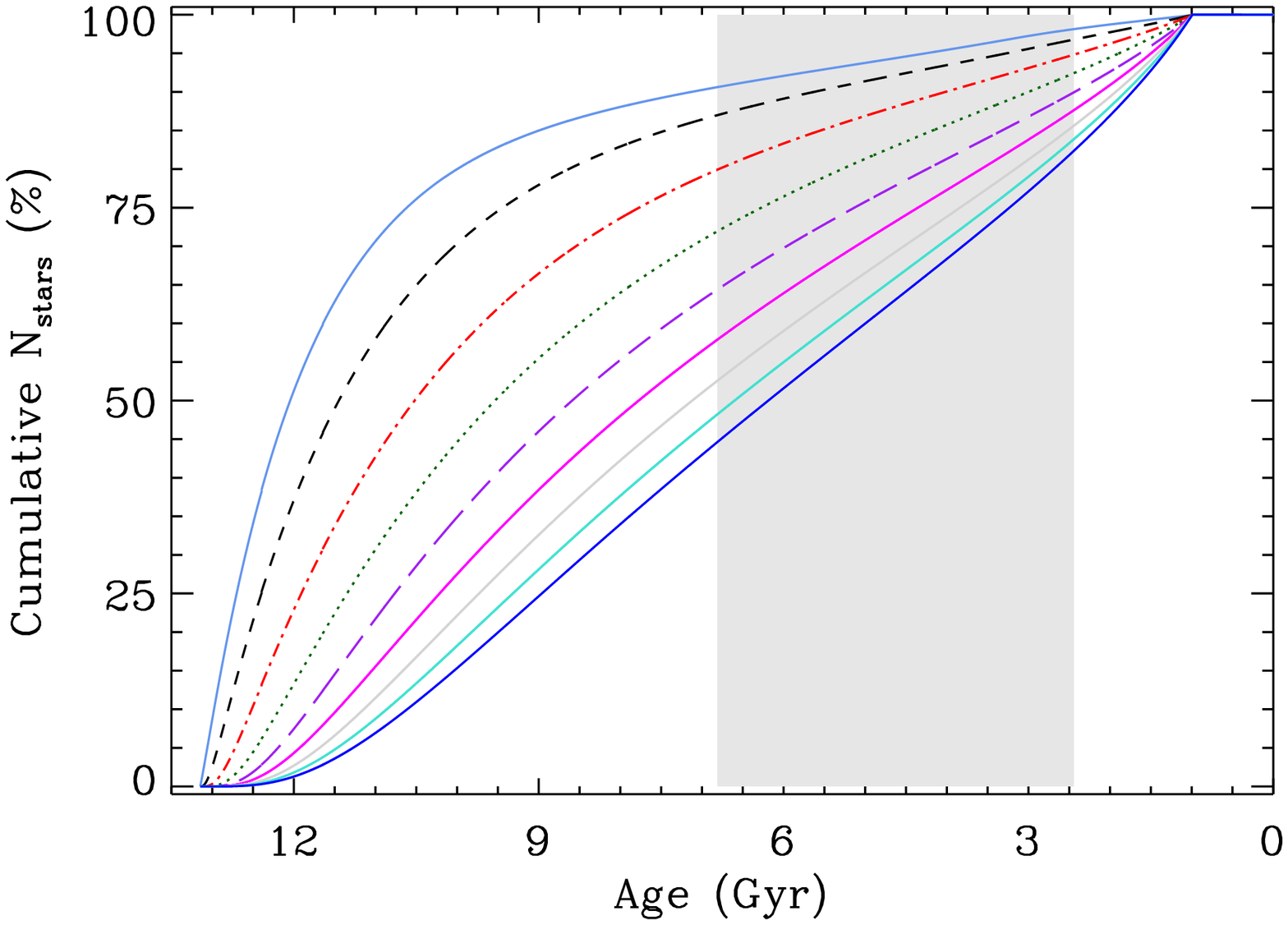}
   \caption{Percentage cumulative number of stars formed as a function of the age in Gyr. The filled-grey region highlights the age interval in which the fraction of bars in the universe is observed to increase. The models are plotted with lines and colours as in the previous figures.}  
    \label{fig:cumulative_stars}
\end{figure}

Fig. \ref{fig:cumulative_stars} shows the percentage cumulative number of stars created as a function of the age in Gyr. As in the previous figures, the age equals to zero corresponds to the present time. The number of stars for each time-step in the model was calculated by using the empirical formula for stellar life-times from \citet{raiteri96} as well as the \citet{kroupa01} IMF. The inner regions of the bulge reach a higher percentage of formed stars at earlier times than the outer regions, as expected from the collapse time-scales for each region. We see that for ages younger than $\sim$6\,Gyr a higher percentage of the stars is formed in the model for all the radial regions. The inner regions also reach a higher percentage of stars formed earlier than the outer ones, because the infall rates in the inner regions are higher. 

For what refers to the age of the Galactic bar, it is not well constrained in the literature. There exist, however, some indications about the time the bar formation and the creation of stars in the regions bar/bulge. For instance, \citet{cole02} indicate an upper limit for the Galactic bar age of 6\,Gyr from its infrared carbon star population. They also suggested that the age of the Galactic bar is probably about 2\,Gyr. On the other hand, \citet{bovy19}, using APOGEE and Gaia DR2 data, analysed the age and stellar abundances of the stars within the bar region and found that the stars in the bar likely formed 8\,Gyr ago, while stars in the bulge outside of the bar are younger and more metal-rich. However, as discussed by \citet{baba20}, the age of the stars in the bar is not necessarily the age of the bar itself, since the bar can capture stars that were formed before the bar formation. By looking for the evolution of the fraction of barred galaxies with redshift in the universe, \citet{sheth08} show that there is a dependence with the redshift and the lower the bar fraction, the higher the redshift. In particular, they find that the bar fraction drops from $\sim 65$\% in the local universe to $\sim 20$\% at $z\sim 0.84$. An independent work from \citet{melvin14} also find that the bar fraction decreases with the increase of the redshift. Their figure 9, which is a compilation of different high-redshift studies, shows that the fraction of barred galaxies increases from $z\sim0.8$ to $z\sim0.4$. They also show that the fraction of bars in massive galaxies with $\log(M/M_{\sun}) > 10.64$ reach a maximum at $z\sim0.5$. Therefore, galaxies as the Milky Way likely formed their bars in the range $0.2< z < 0.84$, although a bar formation at higher $z$ is not excluded. This way, we plotted in Fig. \ref{fig:cumulative_stars} a filled-grey region highlighting the expected age interval for bars formation in the universe, considering the $z$ range above. For a older bar ($\sim$\,6 Gyr) a high percentage ($\gtrsim 75\%$) of the stars were already formed in the inner regions of the bulge (that is, before the bar formation). On the other hand, for a younger bar ($\sim$3\,Gyr) this high percentage is achieved in all regions in the bulge. In this case, most stars were formed in the bulge before the bar creation. The implications for the chemical evolution of the bar/bulge will be discussed in the next section.

\section{Discussion}
\label{discussion}

From the exploitation of different datasets over the last one and half decades, a state-of-the art observational picture of the bulge has emerged: the bulge is a very complex region of the Milky Way, with the potential concurrence of at least two different components/populations, and with observational properties suggesting a complex formation and evolutionary history. The complexity of the bulge MDF was one of the first evidences towards this scenario. There are additional observational evidences beyond the metallicity, highlighting differences between stars in each one of the metallicity components of the bulge MDF:

\begin{enumerate}
\item Kinematically, metal-rich stars are associated with the bar stream motions and general kinematics, as expected from numerical simulations \citep{babusiaux10,ness13,wegg19}. On the other hand, metal-poor stars present a more isotropic and kinematically hot distribution.

\item The double red clump, a feature revealing the underlying X-shape of the buckled bar, has been detected associated to metal-rich stars only \citep{ness12,rojas-arrigada14}.

\item The projection in the ($\ell,b$) plane of the spatial density distribution of metal-rich (boxy) and metal-poor (spherical-like) stars is different, as highlighted by the density maps of \citet{zoccali17}, with each component amounting to about 50\% of the stellar mass \citep{zoccali18}.

\item The age demography of metal-rich and metal-poor components seem to be different: while metal-poor stars are all overall old, metal-rich stars span a wide range of ages, including a fraction of young stars \citep{bensby17,schultheis17}.

\item Metal-rich and metal-poor stars are in different locus of the [$\alpha$/Fe] {\it vs.} [Fe/H] plane. They both display a sequence-like distribution, but recently, based on larger and more precise datasets, it has been found that even there is a gap (a low density region) between them \citep{rojas-arriagada19,queiroz21}.
\end{enumerate}

These observations seem to argue for a dichotomy in the structural, dynamical, and eventually age composition, of metal-poor and metal-rich bulge stars. In light of these evidences, the metal-poor stars seem to have formed in the initial phases of the formation of a classical bulge, while the metal-rich stars would be associated with the bar and an origin related with the secular evolution of the early (thin/kinematically cool) disc. 

The CEM presented in this work can be inserted in this same scenario considering that the origin of the metal-poor and a fraction of the metal-rich stars of the bulge are associated with a spheroidal collapse where the time-scale has a dependency on the radial distance to the central region. The inner regions have shorter collapse time-scales, while the outer regions have longer ones, according to Table \ref{tab:tcoll}. This scenario could explain the differences observed in the MDFs from distinct radial regions of the bulge. However, to better reproduce the observational data, we still need other stars with metallicity higher than [Fe/H] $ > 0.25$\,dex, specially at the inner regions (0.25, 0.5 and 0.75 kpc) in Fig.~\ref{fig:mdf}. 

The formation of the bar may induce radial gas flows in these regions, that can form another population of metal-rich stars, not seen in simulations presented here, since the radial gas flows are not included in the model. This additional metal-rich population could be responsible for the bulge double-peak MDF and the gap (the low density region) in the [$\alpha$/Fe] {\sl vs.} [Fe/H] plane seeing by observations \citep[e.g.][]{queiroz21}. \citet{bensby17} proposed a conservative fraction of a 15\% genuinely metal-rich young stars ([Fe/H] > 0.0 and younger than 5\,Gyr) by taking into consideration their observational bias. This additional formation of metal-rich stars population could be triggered by the formation of the  boxy/peanut X-shape bar bulge in the Milky Way galaxy. As noted by \cite{barbuy18}, the existence of the boxy/peanut bulge, however, is not incompatible with a scenario as the one described in this work. The agreement between the results of the model presented in this work with data, particularly the results of Fig.~\ref{fig:mdf} and \ref{fig:alpha_fe}, seems to indicate that a large fraction of stars in the bulge were likely formed by the mixed scenario discussed above. 

There are also observational evidences based on morphological, photometric and kinematic properties, for the existence of composite bulges (boxy-peanut and classical bulges coexisting in the same galaxy) as e.g. \citep{prugniel01, mendez-abreu14, erwin15, fisher16}, and they point towards a complex formation and evolutionary scenario. The inclusion in the model of radial gas flows of gas resulted from the secularly evolved disc into a boxy/peanut X-shape bar is necessary to explain an additional formation of metal-rich stars as reported by the observations, as e.g. the less good agreement between model and observations for metal-rich stars in Fig.~\ref{fig:alpha_fe}. From Fig.~\ref{fig:mdf}, the inclusion of the radial gas flows and, consequently, the formation of the metal-rich population, are also necessary to increase the number of stars in the metal-rich parts of the MDFs (i.e [Fe/H] > 0.25\,dex), particularly in the inner regions panels: GC, 0.25\,kpc (1.7\degr),  0.50\,kpc (3.5\degr) and 0.7\,5kpc (5.2\degr), where we see that the model is predicting a low number of stars at these metallicities. 

This way, the existence of a bar may explain a metal-rich young stars component. The bar would occur, however, at times later than the maximum star formation occurs in our modelled regions. Following \citet{cole02} the bar is at least 3\,Gyr old, and very probably formed in the last 6\,Gyr. The filled-grey region in Fig.~\ref{fig:cumulative_stars} highlights the expected age interval for bars formation in the universe. If the 15\% genuinely metal-rich young stars are formed in the last $\sim$5\,Gyr, as proposed by \citet{bensby17}, most of the stars of the bulge where already formed, specially at the inner regions. Therefore, it is possible that the most-metal rich stars that appear in our model pre-existed the bar. There are previous simulations that support this possibility of old metal-rich stars; e.g. \citet{debattista17} and \citet{fragkoudi17} find that metal rich stars there exist and were kinematically cold at the time of bar formation and, therefore, easily captured onto bar orbits. The simulations from \citet{aumer15} also suggested that probably bars are formed by capturing young stars that are dynamically cool and formed from gas located outside the bar. 
The scenario proposed here also explains the recent results from \citet{wegg19}, pointing to stars belonging to the  bar of the Milky Way outside the boxy bulge (the so-called long bar) are more metal rich than the inner disc.

We note that the model from \citet{haywood18} reproduces the global MDF and the age-metallicity relation in the bulge by proposing a CEM in a scenario where the bulge is formed in a modified closed-box model and relaxing the instantaneous recycling approximation. The bulge is formed similarly to the inner disc and some observational constraints are reproduced, as e.g. the dip in the MDF at [Fe/H]$\sim0$ dex. This dip was interpreted as the quenching in the SFR due to the transition from the thick to the thin discs. However, their model predicts a high number of low metallicity stars not seeing in the observations and the variations of the MDF in different regions within the bulge were not explored.

Therefore, the model presented in this work has the advantage over previous CEM in the literature of minimising some free parameters as the gas accretion time-scale and the IMF, the last one being the same for the disc and the bulge. We find that an inside-out formation of the bulge is a plausible scenario to reproduce the majority of the chemical properties from the stars pertain to the bulge. The inclusion of the radial gas flows resulted from the boxy/peanut X-shape bar formation is only necessary to explain the formation of an estimated 15\% fraction of genuinely young metal-rich stars population. Nonetheless, the inclusion of the Galactic bar is out of the scope of this study and we will consider this in a forthcoming paper. However, we point that in \citet{cavichia14}, the region of the bulge connected to a disc was simulated by using a model with radial flows. These radial flows increased the star formation and the Oxygen abundances in the region located in the border bulge-disc, but the MDF was not very modified by them. Perhaps the 15\% of metal-rich stars missing from our MDFs in outer regions, could be explained in a future model with radial gas fluxes.

\section{Conclusions}
\label{conclusions}
In this work, we have studied the chemical evolution of the Milky Way bulge by updating our previous models presented in \citet{molla95,molla00,molla05,cavichia14,molla15,molla16,molla19a}. Specifically, we develop a new model in which the bulge is formed inside-out in a multi-zone approach. The S\'ersic brightness profile is used to derive the radial mass distribution inside the bulge, until 2.1\,kpc, which is the adopted bulge radius in the model. The most recent stellar yields for low, intermediate-mass stars from \citet{cristallo11,cristallo15} are adopted. For massive stars, the stellar yields from \citet{limongi18} are used. The  Delay Time Distribution that rules the supernovae type Ia production rate, an additional update from our previous model, for which we take \citet{strolger20} function. Abundance ratios are one of the most useful observational constraints to define the time-scale for the formation of structures in a galaxy. We analysed the stellar MDF, the diagrams  [X/Fe] {\sl vs.} [Fe/H], and the radial abundance distributions, to impose constraints on the history of star formation in the bulge. The model results are compared against the new available data from recent spectroscopic surveys, as e.g. the Gaia-ESO \citep{gilmore12}, APOGEE \citep{majewski17}, GIBS \citep{zoccali17}, and also other observational data available in the literature. We can summarize the most important results as:

\begin{itemize}
    \item The inside-out formation of the bulge is characterized by an infalling gas rate that has a shorter collapse time-scale in the inner regions ($\tau \sim 1$\,Gyr) while at the outer bulge regions will be formed with a longer time-scale ($\tau \sim 4$\,Gyr). 
    
    \item The metallicity distribution function (MDF), represented by the [Fe/H] histograms for each spherical region, as computed for this model, predicts a wider MDF for the inner regions and a narrower for the outer ones. Also, the number of stars composing the metal poor (MP) population is higher in the central fields than the outer ones, in agreement with the data. At the same time, the fraction of metal rich (MR) stars increases towards the centre. Nonetheless, the positions of the peaks do not change considerably. The global MDF is asymmetrical and is reproduced by the model summing the results from all the radial regions. In the model the asymmetry in the MDF seems to originate from the combination of stellar populations with different collapse time-scales. 
    
     \item The data shows a bimodal MDF at the outer regions of the bulge in the border with the disc that is not reproduced by the model. Another scenario is needed to reproduce this feature, as e.g. radial gas inflows to build the box/peanut X-shape bar. This will be considered in a future work. 
 
    \item The model predicts that there is a radial [Fe/H] gradient of -0.12\,dex\,kpc$^{-1}$ within the bulge. The spread of the stellar data in the [Fe/H] {\it vs.} $R$ plane do not allow us to draw firm conclusions and more accurate data is necessary. 
     
    \item The O/H radial distribution is compared with data from planetary nebulae (PNe). The model predicts a negative gradient for $R < 2$\,kpc, and approximately flat at the very inner regions $R < 0.25$\,kpc. Between $2< R < 3$\,kpc there is a step in the radial distribution, which is seen in both data and model. However, more quality data is needed to confirm this trend. 
    
    \item In the [$\alpha$/Fe] {\sl vs.} [Fe/H] plane, where $\alpha$ stands for O, Mg, Si and Ca, the model predicts a dependence with the spherical region. The inner bulge regions have higher [$\alpha$/Fe] for subsolar metallicities. At higher metallicities the differences are less important. The results suggest that the spread of the data in this kind of plot might be the result of mixing stars from different radial regions. The differences are in the range of $\sim 0.04$ -- $0.08$\,dex, doing very difficult to detect them observationally considering the current uncertainties in the observational data. Nonetheless, there are observational results in the literature pointing to a dispersion in the same order of magnitude.
    
    \item The [Fe/H] age evolution for each radial region is in agreement with the observational data and suggests an age-metallicity relation: subsolar metallicity stars have a narrower range of ages and are mostly old; supersolar metallicity stars span a very wide range of ages. There is a slight dependence of [$\alpha$/Fe] ratio with age and stars with high [$\alpha$/Fe] tend to be older.
    
    \item The agreement between the
results of the model presented in this work with data seems to indicate that a large fraction of stars in the bulge were likely formed by a mixed scenario. First the collapse of a spheroidal component with different collapse time-scales, then the formation of a boxy/peanut X-shape bar when mostly ($\sim 75\%$) of the stars were already formed in the bulge. The inclusion of the radial gas flows resulted from the boxy/peanut X-shape bar formation (not performed here) is only necessary to explain the formation of only an estimated 15\% fraction of genuinely young metal-rich stars population.

\end{itemize}

\section*{Data availability}

The data underlying this article will be shared on reasonable request to the corresponding author.

\section*{Acknowledgements}

The authors wish to acknowledge the anonymous referee, whose comments improved the quality of this paper. This work has made use of the computing facilities available at the Laboratory of Computational Astrophysics of the Universidade Federal de Itajub\'a (LAC-UNIFEI). The LAC-UNIFEI is maintained with grants from Capes, Cnpq and FAPEMIG. O.C. would like to thank CAPES, PGF/UNIFEI and Funda{\c c}\~ao de Amparo \`a Pesquisa do Estado de Minas Gerais (FAPEMIG) grant APQ-00915-18. This work was partially supported by Spanish grants from the former MINECO-FEDER AYA2016-79724-C4-3-P and PID2019-107408GB-C41 funded by MCIN/AEI/10.13039/501100011033. 

%%%%%%%%%%%%%%%%%%%%%%%%%%%%%%%%%%%%%%%%%%%%%%%%%%

%%%%%%%%%%%%%%%%%%%% REFERENCES %%%%%%%%%%%%%%%%%%

% The best way to enter references is to use BibTeX:

\bibliographystyle{mnras}
\bibliography{biblio} 

% Don't change these lines
\bsp	% typesetting comment
\label{lastpage}
\end{document}